\documentclass[a4paper,11pt]{article}
\pdfoutput=1 

\usepackage{jheppub} 

\usepackage{amsmath,amssymb,mathrsfs,mathtools}

\def\cA{\mathcal{A}}
\def\cK{\mathcal{K}}

\def\cW{\mathcal{W}}

\def\zt{{\bar z}}
\def\Li{{\rm Li}}
\def\ep{\epsilon}
\def\cO{{\cal O}}
\def\cN{{\cal N}}
\def\ra{\rightarrow}
\def\Neq#1{$\mathcal{N}{=}\,#1$}
\newcommand{\be}{\begin{equation}}
\newcommand{\ee}{\end{equation}}
\def\beq#1\eeq{\begin{align}#1\end{align}}
\newcommand{\nn}{\nonumber}

\title{\boldmath Perturbative solutions of ${\cal N}=1^*$ holography on $S^4$}

\author[a,b]{Nakwoo Kim,}
\author[a,c]{Se-Jin Kim}

\affiliation[a]{Department of Physics and Research Institute of Basic Science,
	Kyung Hee University,\\Seoul 02447, Republic of Korea}
\affiliation[b]{School of Physics, Korea Institute for Advanced Study, \\Seoul 02445, Republic of Korea}
\affiliation[c]{Theoretical Physics Department, CERN, CH-1211 Geneva 23, Switzerland}

\emailAdd{nkim@khu.ac.kr}
\emailAdd{sejin.kim@cern.ch}

\abstract{We apply the recently proposed perturbative technique to solve the supergravity BPS equations of ${\cal N}=1^*$ theories put on $S^4$. In particular, we have calculated the coefficients of the leading quartic terms exactly, in the expression of the universal part for the holographic free energy as a function of the mass parameters. We also report on the coefficients of higher order terms up to 10th order, which are computed numerically.}

\begin{document} 
\maketitle
\flushbottom

\section{Introduction}
\label{sec:intro}
According to AdS/CFT correspondence \cite{Maldacena:1997re}, we have a large number of duality pairs where we have on one side strongly coupled gauge field theories and on the other side supergravity solutions with an anti-de Sitter factor in string/M-theory. Thanks to the breakthrough made in \cite{Pestun:2007rz}, using the supersymmetric localization technique, at least for theories with enough number of supersymmetries and appropriate coupling to background geometry, one can reduce a certain class of path integrals to a finite-dimensional matrix integral. When the field theory does have a holographic dual, the large-$N$ limit of the matrix integral should match the counterpart quantities, {\it e.g.} Wilson loops and free energy, in supergravity. This program has been successfully applied to many examples in various dimensions, see for instance \cite{Zarembo:2016bbk} for more details and references.

In this paper we are interested in non-conformal deformations of AdS/CFT. Within the lower-dimensional supergravity theories, which are related to 10/11 dimensional supergravity through a consistent truncation, we have scalar fields which are dual to relevant and marginal operators in the dual theory. In particular, those scalars which are dual to mass terms can be identified and the associated BPS flow equations can be written down. These BPS equations are first-order nonlinear differential equations, and exact solutions are usually not available. When one evaluates the holographically renormalized action for regular solutions, it is expected to match the large-$N$ limit of mass-deformed free energy for the dual field theory. 
This task is sometimes called ``precision holography'' for mass-deformed conformal field theories. It started in \cite{Freedman:2013ryh}, where exact supergravity BPS solutions to mass-deformed ABJM model \cite{Aharony:2008ug} were constructed and its renormalized supergravity action was shown to agree with the large-$N$ limit of localization computation with supersymmetric mass terms. Then the authors of  \cite{Bobev:2013cja} tackled a similar problem of comparing the so-called $\cN=2^*$ super Yang-Mills theory in $D=4$ (which can be obtained by giving mass to $\cN=4$ super Yang-Mills theory in a way compatible with $\cN=2$ supersymmetry), and their supergravity dual solutions. By analyzing the numerically constructed supergravity solutions, it was argued in \cite{Bobev:2013cja} that the evaluated action matches the prediction of localization computation. See also \cite{Bobev:2018hbq} for the uplift of the mass-deformed supergravity solutions from $D=5$ to $D=10$ and its application.

In the construction of supergravity solutions and their holographic analysis, the crucial information we need is a relation between the integration constants in the UV expansion, constrained by regularity at IR. In earlier works like \cite{Bobev:2013cja,Bobev:2016nua,Gutperle:2018axv}, such relations were inferred from numerical solutions. It is certainly not satisfactory, especially when analytic expressions are available from the localization computations. In a recent work \cite{Kim:2019feb}, one of the present authors proposed a perturbative method using which one can analytically extract the aforementioned relation in a series expansion form. In \cite{Kim:2019feb}, three non-conformal holography problems were addressed: mass-deformed ABJM, the mass deformation of Brandhuber-Oz superconformal field theory in $D=5$ \cite{Brandhuber:1999np}, and also the $\cN=2^*$ deformation of $\cN=4, D=4$ super Yang-Mills.
For the first two examples, one can either find exact supergravity solutions \cite{Freedman:2013ryh}, or at least the series form of the relation between integration constants can be summed \cite{Kim:2019feb}. On the other hand, for the case of mass-deformed $AdS_5$, the linearized equations involve log and polylogarithm terms which render explicit integrations difficult, and \cite{Kim:2019feb} resorted to approximation using series expansion up to high but finite orders. It is the purpose of this paper to improve and extend the study of mass-deformed $\cN=4, D=4$ super Yang-Mills in the gravity dual side. In particular, we tackle the problem of $\cN=1$ mass deformations, which is also called $\cN=1^*$ models. The dual gravity action was already constructed as a consistent truncation of $D=5$ maximal $SO(6)$-gauged supergravity by Bobev {\it et al.} in \cite{Bobev:2016nua}. Combining both brute-force integration and also Pad\'e approximants, we report a series expansion form of gravity side free energy, containing two more coefficients than the numerical results of \cite{Bobev:2016nua}. We also push the integration analytically to 3rd order, while in \cite{Kim:2019feb} we used approximation already at 3rd order.

The plan of this report is as follows. In Sec.\ref{sec:2} we present the five-dimensional Einstein-scalar action we are interested in, and also its associated BPS equations. In Sec.\ref{general}, we describe the general feature of our perturbative approach when applied to holographic $\cN=1^*$ system. We also report on the analytic results up to 3rd-order perturbation, for general three-mass case. In Sec.\ref{singlem} we consider three special cases of single mass models, and report on the expression of holographic free energy obtained from perturbation up to 9th order. We conclude with discussions in Sec.\ref{discuss}.

\section{Summary of previous works}
\label{sec:2}
The aim of this paper is to re-visit the analysis of BPS equations presented in \cite{Bobev:2016nua}. The action derived there is for a five-dimensional Einstein gravity which additionally contains ten (real)\footnote{Originally in Minkowski signature they are four complex scalar fields $z_a$ and two real scalars $\eta_1,\eta_2$, but in Euclidean signature $z_a,{\bar z}_{\bar a}$ are treated independent. For our purpose it is fine to treat them simply as 10 real scalars.} scalar fields. In terms of AdS/CFT, those scalars are dual to various operators appearing in the action of \Neq4 super Yang-Mills theory when we put the theory on $S^4$ and turn on mass terms of three chiral multiplets and keep \Neq1 supersymmetry. In this article we will closely follow and use the formulae in \cite{Bobev:2016nua}, with minor notational changes.\footnote{For instance, $\exp(\beta_i^{there})=1/\eta_i^{here}$.}
Thanks to supersymmetry, the Lagrangian density can be written succinctly in the following manner. 
\beq
{\cal L}&=-\frac{1}{4}R + 3 \frac{(\partial{\eta_1})^2}{\eta_1^2}+\frac{(\partial{\eta_2})^2}{\eta_2^2} + \frac{1}{2}{\cal K}_{a \bar{b}} \partial_{\mu} z^a \partial^{\mu} \bar{z}^{\bar{b}} -{\cal P} ,
\\
{\cal P}& = \frac{1}{8} e^{\cal K} \left(\frac{\eta_1^2}{6} \partial_{\eta_1}{\cal W} \partial_{\eta_1}{\cal \widetilde{W}}+\frac{\eta_2^2}{2} \partial_{\eta_2}{\cal W}\partial_{\eta_2}{\cal \widetilde{W}}+{\cal K}^{\bar{b} a}\nabla_a {\cal W} \nabla_{\bar{b}} {\cal \widetilde{W}}-\frac{8}{3} {\cal W} {\cal \widetilde{W}}\right) ,
\\
{\cal W} &=
\eta_1^{-2}\eta_2^{-2}\left(1+z_1z_2+z_1z_3+z_1z_4+z_2z_3+z_2z_4+z_3z_4+z_1z_2z_3z_4\right)
\nn\\
&+
\eta_1^{-2}\eta_2^2\left(1-z_1z_2+z_1z_3-z_1z_4-z_2z_3+z_2z_4-z_3z_4+z_1z_2z_3z_4\right)
\nn\\
&+
\eta_1^4\left(1+z_1z_2-z_1z_3-z_1z_4-z_2z_3-z_2z_4+z_3z_4+z_1z_2z_3z_4\right),
\eeq
and ${\cal K}= - \sum_{a=1}^4 \log(1-z^a \bar{z}^{\bar{a}})$, ${\cal K}_{a \bar{b}} \equiv \frac{\partial^2 {\cal K}}{\partial z^a \partial \bar{z}^{\bar{b}}}$. ${\cal \widetilde{W}}$ is the same as ${\cal W}$, except for the replacement of $z^a$ by $\bar{z}^{\bar{a}}$.

Note that here $\eta_1,\eta_2$ are two real scalar fields, while $z_a,{\bar z}_b \;(a,\bar{b}=1,2,3,4)$ together originally constitute four complex scalars. Since we consider supergravity in Euclidean signature, $z_a$ and ${\bar z}_{\bar a}$ are not mutually conjugate any more and we treat them as independent real scalars. We choose conformal gauge for the metric convention, 
\be
\label{confgauge}
ds^2 = e^{2A} (dr^2/r^2 + ds^2(S^4)) . 
\ee
Then the AdS vacuum solution has unit radius, with the following scalars and warp factor
\beq
e^{2A} = \frac{4r^2}{(1-r^2)^2} , \quad
 \eta_1 = \eta_2 = 1 , \quad
 z_a = \zt_{\bar b} = 0 .
\eeq

The superpotential $\cW$ and the K\"ahler potential $\cK$ carry the information of supersymmetry transformations and eventually determine the BPS equations which provide first-order differential relations for the scalar fields and warp factor. In the conformal gauge Eq.\eqref{confgauge}, the BPS equations can be written as follows.\footnote{These equations are equivalent to Eq.(4.14) in \cite{Bobev:2016nua}. Here they are re-arranged to better suit our perturbative prescription.
}
\beq
\begin{aligned}
 \partial _rz^a&=-\frac{3}{2}(\partial_r A \pm 1/r){\cal K}^{a \bar b} \frac{\partial}{\partial \bar z ^b}\log ({\cal W \widetilde{W} }e^{\cal{K}}) 
 ,
\\ 
 \partial _r \bar{z}^{\bar b}&=-\frac{3}{2}(\partial_r A \mp 1/r){\cal K}^{a \bar b} \frac{\partial}{\partial z ^a} \log ({\cal W \widetilde{W} }e^{\cal{K}}) ,
\\
\partial_r \eta_1 &= -\frac{\eta_1^2}{72}\frac{e^{2A}}{r^2 \partial_r A}\frac{\partial}{\partial \eta_1} ({\cal W \widetilde{W} }e^{\cal{K}}) , 
\\
 \partial_r \eta_2& = -\frac{\eta_2^2}{24}\frac{e^{2A}}{r^2 \partial_r A}\frac{\partial}{\partial \eta_2} ({\cal W \widetilde{W} }e^{\cal{K}}) ,
\\
(\partial_r A)^2&=\frac{1}{r^2}+\frac{1}{9}\frac{e^{2A}}{r^2} ({\cal W \widetilde{W} }e^{\cal{K}}) 
,
\\
\frac{\partial}{\partial \eta_i} {\cal W }&=\frac{\partial_r A\pm 1/r}{\partial_r A\mp 1/r} \frac{{\cal W}}{{\cal \widetilde{W}}}\frac{\partial}{\partial \eta_i} {\cal \widetilde{W}} .
\end{aligned}
\eeq

One may try to solve these equations near $r=1$ (UV). Then it turns out that generally the solutions contain eight integration constants. For more details readers are referred to eq.(4.19) in \cite{Bobev:2016nua}, and also \eqref{muv} of this paper. In terms of $\rho=2\tanh^{-1} r$, the UV expansion ($\rho\ra\infty$) contains
\beq
(z_1 +z_2 +z_3+z_4 + \bar z_1 + \bar z_2+ \bar z_3+ \bar z_4)/4 &=  \left(1-s^2\right) \left(2 \mu _1 \rho+v_1-s \mu _2 \mu _3 \right) e^{-2\rho} + \cO (\rho^2e^{-4\rho}),
\nn\\
(z_1 -z_2 +z_3-z_4 + \bar z_1 - \bar z_2+ \bar z_3- \bar z_4)/4 &=  \left(1-s^2\right) \left(2 \mu _2 \rho+v_2-s\mu _1 \mu _3 \right) e^{-2\rho} + \cO (\rho^2e^{-4\rho}),
\nn\\
(z_1 +z_2 -z_3-z_4 + \bar z_1 + \bar z_2- \bar z_3- \bar z_4)/4 &= \left(1-s^2\right) \left(2 \mu _3 \rho+v_3-s\mu _1 \mu _2 \right)e^{-2\rho} + \cO (\rho^2e^{-4\rho}),
\nn\\
(z_1 -z_2 -z_3+z_4 + \bar z_1 - \bar z_2- \bar z_3+ \bar z_4)/4 &= 2s-\frac{s}{2} \left(1-s^2\right) \left(\mu _1^2+\mu _2^2+\mu _3^2\right) e^{-2\rho}+{\cal O}(\rho e^{-4\rho}), 
\nn\\
(z_1 -z_2 -z_3+z_4 - \bar z_1 + \bar z_2+ \bar z_3- \bar z_4)/4 &=
- \frac{1}{2} \left(1-s^2\right)\Big[ 2w-\left(1-3 s^2\right)\mu _1 \mu _2 \mu _3
\nn\\
\MoveEqLeft[12] 
-2 s \left(\mu _1 v_1+\mu _2 v_2+\mu _3 v_3\right)-4s \left(\mu _1^2+\mu _2^2+\mu _3^2\right)  \rho \Big] e^{-3\rho}+{\cal O}(\rho e^{-5\rho}). 
\label{uv:gen}
\eeq
As explained in \cite{Bobev:2016nua}, $\mu_1,\mu_2,\mu_3$ are interpreted as sources for the mass of three chiral multiplets in the field theory, $v_1,v_2,v_3$ are the expectation values of mass term operators, $w$ is dual to gaugino expectation value, and $s$ is to be identified with the Yang-Mills coupling constant. We also note that, to be precise the holographic dictionary identifies $\mu_i=\pm i m_i a$ \cite{Bobev:2013cja,Bobev:2016nua}, where $m_i$ is the mass of chiral multiplets and $a$ is the radius of $S^4$ where the gauge field theory is put on. The crucial information we need is how $v_i$ are determined as functions of $\mu_i$, once we demand IR regularity ($r=0$). More specifically, according to the holographic computation in \cite{Bobev:2016nua}, the holographic free energy $F$ satisfies
\be
\frac{\partial^3 F}{\partial \mu^3_i} = -\frac{N^2}{2} \frac{\partial^2 v_i}{\partial \mu^2_i} . 
\ee
It turns out that the localization side computation for $F$ contains a scheme-dependent factor, which can be removed when we take third or higher derivative. Integration of the above expression with $F=F'=F''=0$ is thus called the {\it universal} part. In this paper from now on we will always mean the universal part, when we refer to $F$. 
\section{Perturbation for general solutions}
\label{general}
In our perturbative approach we take the Euclidean AdS (i.e. hyperbolic space) as a zeroth-order reference solution. At first, we turn on scalars $z_a,{\bar z}_{\bar b}$, while $A,\eta$ are still at vacuum configuration. Then at the second order, through gravitational and inter-scalar interactions we begin to have non-vacuum values for warp factor $A$ and scalar field $\eta$. It turns out that via appropriate choice of the perturbative parameter $\ep$, without losing generality we may set
\beq
\begin{aligned}
z_a(r)& = \sum _{n =0}^{\infty} \epsilon^{2n+1}z_a^{(2n+1)}(r)  , \quad \quad \bar{z}_{\bar b} (r)= \sum _{n=0}^{\infty}\epsilon^{2n+1} z_{\bar b}^{(2n+1)} (r) ,
\\
\eta_i(r) &= 1+ \sum_{n=1}^{\infty} \epsilon^{2n} \eta_i^{(2n)}(r)   \quad  \ (i= 1, 2),
\\
e^{2A(r)}&=\frac{4 r^2}{(1-r^2)^2} \left( 1 +\frac{1+r^2}{1-r^2}\sum_{n=1}^{\infty} \epsilon^{2n}{\cal A}^{(2n)} (r) \right). 
\end{aligned}
\eeq
We substitute this perturbative expansion into the BPS equations and choose the upper sign for concreteness. Demanding that the equations should be satisfied for arbitrary $\ep$, and doing some algebraic manipulation, one obtains the following form of equations.
\beq
\label{peq}
\begin{aligned}
\partial _rz_a^{(2n-1)}&=-\frac{1}{r(1-r^2)}\Xi_a^{(2n-1)} , \quad \partial _r\bar{z}_{\bar{b}}^{(2n-1)}=-\frac{r}{1-r^2}\widetilde{\Xi}_{\bar b}^{(2n-1)} , 
  \\
\eta^{(2n)} &= H^{(2n)}  , \quad \quad 
\partial_r {\cal A}^{(2n)}=-\frac{4r}{3(1+r^2)^2}\Sigma^{(2n)} .
  \end{aligned}
\eeq
The right-hand-side expressions here contain some rational functions of $r$, and also functions  $z_a^{(k)},\zt_{\bar b}^{(k)},\eta^{(k)},{\cal A}^{(k)}$ of degrees $k$ up to $2n-1$. We note that
these equations are always homogeneous: if we assign weight $k$ to functions $z_a^{(k)},\zt_{\bar b}^{(k)},\eta^{(k)},{\cal A}^{(k)}$, the expressions $\Xi^{(k)},{\widetilde \Xi}^{(k)}, H^{(k)}, \Sigma^{(k)}$ also carry weight $k$. Crucially, $\Xi_a^{(2n-1)},{\widetilde \Xi}_{\bar b}^{(2n-1)}$ are in fact linear in $z_a^{(2n-1)},\zt_{\bar b}^{(2n-1)}$ and the in-homogeneous parts are known functions determined from lower orders of $\ep$. Solving the first two coupled differential equations is thus in principle straightforward. Then this result can be substituted into $H^{(2n)}$, determining $\eta^{(2n)}$ algebraically. $\cA^{(2n)}$ is determined through integration of $\Sigma^{(2n)}$, which does not contain $\cA^{(2n)}$ and the rest of perturbative functions are determined already from previous steps. This way, we can determine the solution iteratively to any higher orders, in principle. When we perform the integration, a guiding principle is that the scalar fields should vanish at $r=1$ (in the UV) because we want the solution should be asymptotically AdS. We also demand all functions are regular ({\it i.e.} non-divergent) at $r=0$ (in the IR). 

Let us now illustrate how this scenario leads to explicit solutions. At leading non-trivial order, $n=1$, and we obtain
\beq
\begin{aligned}
&\Xi_1^{(1)}=  
3 z_1^{(1)}+\bar{z}_2^{(1)}+\bar{z}_3^{(1)}-\bar{z}_4^{(1)}
, 
\\
&\Xi_2^{(1)}=  
3 z_2^{(1)}+\bar{z}_1^{(1)}-\bar{z}_3^{(1)}+\bar{z}_4^{(1)}
,
\\
&\Xi_3^{(1)}=  
3 z_3^{(1)}+\bar{z}_1^{(1)}-\bar{z}_2^{(1)}+\bar{z}_4^{(1)}
,
\\
&\Xi_4^{(1)} =  
3 z_4^{(1)}-\bar{z}_1^{(1)}+\bar{z}_2^{(1)}+\bar{z}_3^{(1)}
.
\end{aligned}
\eeq
For $\widetilde{\Xi}^{(1)}$ the expression is essentially the same, except for exchange of $z^{(1)}$ vs. $\zt^{(1)}$ etc.
We see that at this order we just have a coupled system of linear differential equations.  
General solutions should contain eight integration constants, and one can write them as follows
\beq
\begin{aligned}
z_1^{(1)}&=c_1 u_1+\tilde c_5 u_3+c_5+\tilde c_1u_5/r, 
&\bar z_1^{(1)}&=c_1 u_2-\tilde c_5 u_4+c_5+\tilde c_1 r u_5, 
\\
z_2^{(1)}&=c_2 u_1-\tilde c_5 u_3 -c_5+\tilde c_2 u_5/r,
&\bar z_2^{(1)}&=c_2 u_2 +\tilde c_5 u_4-c_5+\tilde c_2 r u_5 ,
\\
z_3^{(1)}&=c_3 u_1-\tilde c_5 u_3-c_5+\tilde c_3 u_5/r,
&\bar z_3^{(1)}&=c_3 u_2+\tilde c_5 u_4-c_5+\tilde c_3 r u_5,
\\
z_4^{(1)}&=c_4 u_1+\tilde c_5 u_3+c_5+\tilde c_4 u_5/r, 
& \bar z_4^{(1)}&=c_4 u_2-\tilde c_5 u_4+c_5+\tilde c_4 r u_5,
\end{aligned}
\eeq
where the constants $c_n$ satisfy $c_4=-c_1+c_2+c_3$ and $\tilde c_4=-\tilde c_1+\tilde c_2+\tilde c_3$. The homogeneous solutions are given as follows.
\beq
u_1(r)&= +\left(1-r^2\right) \left(r-\left(1-r^2\right) \tanh ^{-1}(r)\right)/(2 r^3), 
\nn\\
 u_2(r)&= -\left(1-r^2\right) \left(r+\left(1-r^2\right) \tanh ^{-1}(r)\right)/(2 r),
\nn\\
u_3(r)&= +\left(1-6 r^2-3 r^4\right)/r^3, 
\nn\\
u_4(r) &= + \left(3+6 r^2-r^4\right)/r,
\nn\\
u_5(r)&=+\left(1-2 r^2+r^4\right)/r^2.
\eeq

Due to regularity at $r=0$ we need to set $\tilde c_n=0$. $c_5$ denotes a zero mode, and together with $c_{1},c_2,c_3$ we have four integration constants. In terms of gauge theory language, $c_1,c_2,c_3$ are related to the mass parameter of three chiral multiplets, and $c_5$ is dual to gauge coupling.
We now compare the behavior of our $\cO(\ep)$ solutions to the UV expansion in \cite{Bobev:2016nua}, where the integration constants from UV expansion are called $\mu_1,\mu_2,\mu_3,s$. 
Rescaling $c_n$ to absorb away $\ep$, we find
\beq
\begin{aligned}
 c_1=& - \left(\mu _1+\mu _2+\mu _3\right)/4, 
 \\
 c_2=&- \left(\mu _1-\mu _2+\mu _3\right)/4,
\\
c_3=&-\left(\mu _1+\mu _2-\mu _3\right)/4, 
\\
c_4=&-\left(\mu _1-\mu _2-\mu _3\right)/4, 
\\
c_5=&s.
\end{aligned}
\eeq
And we also obtain that up to this leading order $v_i =- 2 \mu_i$, which is in agreement with the claim in \cite{Bobev:2016nua,Bobev:2018hbq}.

The corrections to warp factor start to appear at $\cO(\ep^2)$. Substituting our $\cO(\ep)$ solutions, we obtain 
\beq
{\cal A}^{(2)'}(r)= \left(\mu _1^2+\mu _2^2+\mu _3^2\right) \frac{r\left(u_1^2-6 u_1 u_2+u_2^2\right)}{6 \left(1+r^2\right)^2} . 
\eeq
We need to integrate it with boundary condition ${\cal A}(r=1)=0$. The result is rather messy, containing polylogarithms. Explicit result can be found in the appendix.

Then the real scalars $\eta_1, \eta_2$ are determined algebraically, and 
\beq
 \eta _{1}^{(2)}(r)&= -\frac{\left(\mu _1^2+\mu _2^2-2 \mu _3^2\right) \left(r^2 u_1^2-u_2^2\right)}{24 \left(1-r^2\right)},
\\
\eta_{2}^{(2)}(r)&=- \frac{\left(\mu _1^2-\mu _2^2\right) \left(r^2 u_1^2-u_2^2\right)}{8 \left(1-r^2\right)}.
\eeq
Again their explicit forms are relegated to the appendix.

At $\cO(\ep^3)$, we substitute our $\cO(\ep^2)$ solutions which play a role of in-homogeneous terms which we need to integrate. Although it is not impossible, the results are quite messy and we do not try to present the result in full detail in this paper. However their precise UV asymptotic behavior can be more easily studied, as the derivative of $\cO(\ep^3)$ solution is determined by $\cO(\ep^2)$ data through BPS equations. We have managed to determine $v_i$ as functions of $\mu_i$. 
The result is as follows, 
\beq
\label{anres}
v_i&=-2 \mu _i+\left(\frac{16 \pi ^4}{525}-\frac{1}{5}\right)  \mu _i^3+\left(\frac{3}{5}-\frac{8 \pi ^4}{525}\right)  \mu _i\left(\mu_1^2+\mu _2^2+\mu _3^2\right) +{\cal O}(\mu_i^5) . 
\eeq

We can also calculate the gaugino condensate holographically, using our solutions. According to the analysis of \cite{Bobev:2016nua}, it is given by $w(\mu)$ which appears in the last line of UV expansion in \eqref{uv:gen}. 
Our $\cO(\ep^3)$ result gives 
\beq
w = 2 \mu_1 \mu_2 \mu_3 +{\cal O}(\mu_i^5) ,
\eeq
which is in agreement with the conjecture made in \cite{Bobev:2016nua}, based on numerical solutions.
\section{Further analysis of single mass models}
\label{singlem}
Now we specialize to three special sub-sectors of the general $\cN=1^*$ models, following \cite{Bobev:2016nua}. On the gauge theory side, we consider first the $\cN=2^*$ theory where we make a hypermultiplet massive in super Yang-Mills theory. Then we also consider two special cases of $\cN=1^*$ deformations. 
\subsection{${\cal N}=2^*$ model}
Compared to the undeformed $\cN=4$ super Yang-Mills, here we give the same non-zero mass to two chiral multiplets ({\it i.e.} a hypermultiplet). The relevant supergravity truncation was constructed earlier in \cite{Bobev:2013cja}.  If we start with the BPS system of the general $\cN=1^*$ cases, we set $z_2=z_4=0$, $\zt_2=\zt_4=0$, $z:= z_1=z_3$, $\zt:=\zt_1=\zt_3$, $\eta_1:=\eta$, $\eta_2=1$. Then the action simplifies to 
\beq
{\cal L}&=-\frac{1}{4}R + 3 \frac{(\partial{\eta})^2}{\eta^2} + \frac{ \partial_{\mu} z \partial^{\mu} \bar{z}}{(1-z \bar{z})^2} +{\cal P} , 
\\
{\cal P}& =- \frac{1}{\eta^4}+2\eta^2\frac{z \bar{z} +1}{z \bar{z} -1}-\frac{\eta^8}{4}\frac{(z-\bar{z})^2}{(z\bar{z}-1)^2} . 
\eeq

We recall the expansion of the solutions near UV, done in the Fefferman-Graham coordinate $\rho=2\tanh^{-1} r$. In particular, the parameters $v,\mu$ are defined in terms of the $\rho\ra\infty$ asymptotic behavior of scalar fields.
\beq
\begin{aligned}
 \frac{1}{2}(z+\zt) &=(v+2 \mu \rho) e^{-2 \rho}
 -(v \mu^2 +2 \mu^3 \rho) e^{-4\rho} 
 + \cdots , 
 \\
 \frac{1}{2}(z-\zt) &=- \mu e^{-\rho}+\frac{\mu^3}{3}  e^{-3\rho}
 + \left(v^2 \mu +4  v \mu^2\rho +4  \mu^3\rho^2 -\frac{2}{15}\mu^5 \right)e^{-5\rho} 
 + \cdots.
 \end{aligned}
\eeq
These are again just the consequence of BPS equations, before imposing the IR ($r=0$) regularity. Based on numerical solutions, the authors of \cite{Bobev:2013cja} conjectured 
\be
v = -2\mu -\mu \log (1-\mu^2) , 
\label{n2con}
\ee
which agrees exactly with the large-$N$ limit of localization computation \cite{Buchel:2013id}. 
This problem was re-visited in \cite{Kim:2019feb} using the perturbative technique we employ in this paper. We note that our refined result up to $\cO(\ep^3)$ in Eq.\eqref{anres} is consistent with the above formula, when {\it e.g.} we set $\mu_3=0,\, \mu:=\mu_1=\mu_2$ it reduces to $v_1=v_2=-2\mu+\mu^3+\cdots$.

Approximation using a truncated series expansion at IR $(r=0)$, to solve the BPS equations, was reported already in \cite{Kim:2019feb}. For the results reported in this paper, we have used an improved method: the BPS equations are solved by series expansion up to certain order, and then the remaining parts are replaced by Pad\'e approximation for substitution to higher $\ep$-order equations. We have performed the computation up to $\cO(\ep^9)$, and find that the result is 
\beq
v(\mu) =-2 \mu+1.00017 \mu ^3 +0.500022 \mu ^5 +0.333344 \mu ^7 +0.250378 \mu ^9 + \cdots.
\label{n2pade}
\eeq
It is confirmed that our perturbative solution agrees nicely with Eq.\eqref{n2con}.

\begin{figure}[h]
\includegraphics[width=0.45 \textwidth]{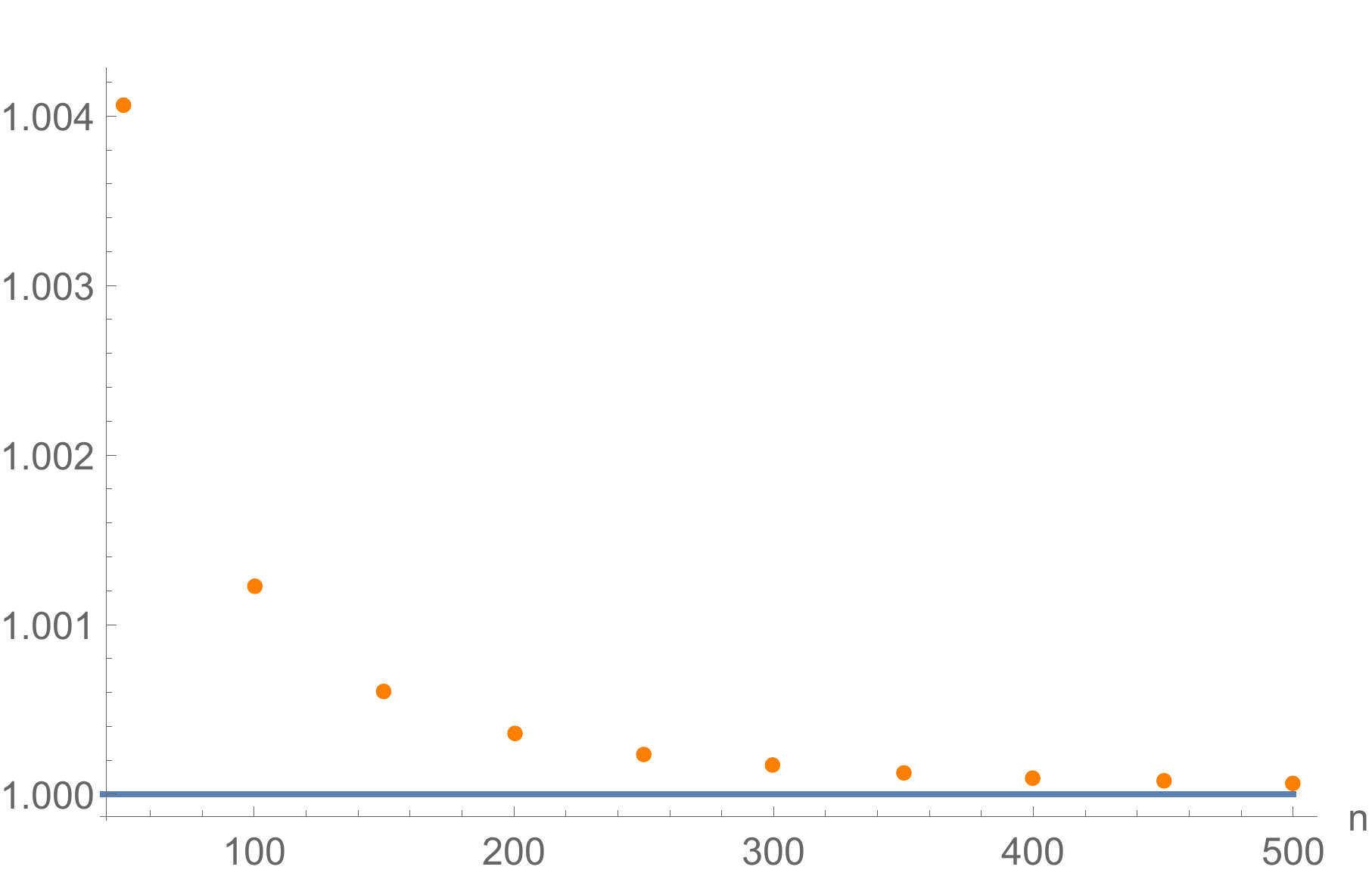} \hskip 1cm
\includegraphics[width=0.45 \linewidth]{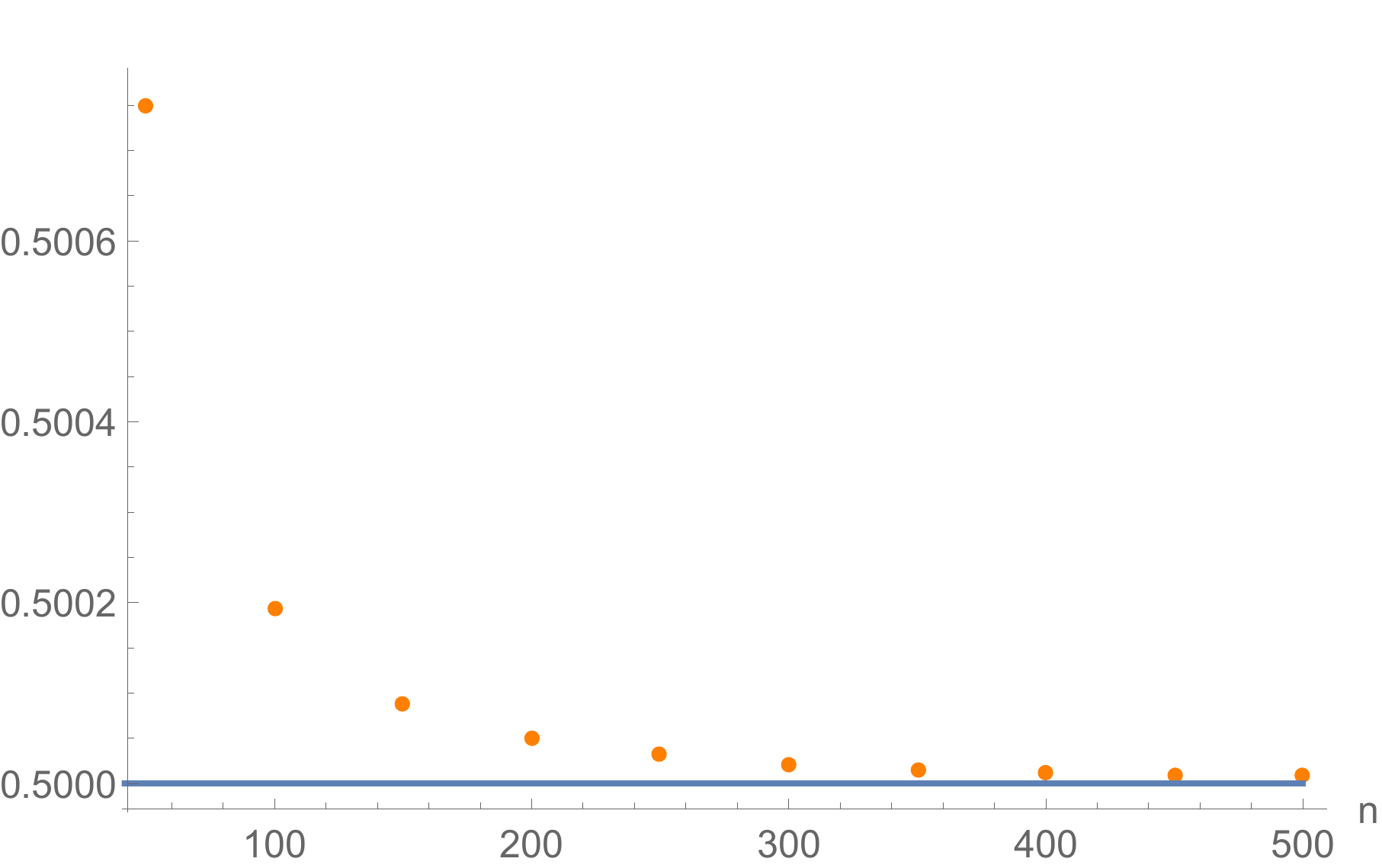} 
  \caption{Plots for the coefficient of $\mu^3,\mu^5$ from the series expansion with Pad\'e approximation, as one increases the truncated length of the expansion.}
  \label{fig1}
\end{figure}

In order to illustrate the reliability of our method, in Figure \ref{fig1} we show how the values of the expansion coefficients of $\mu^3,\mu^5$ in Eq.\eqref{n2pade}, extracted from the limiting behavior of $z_1^{(3)},z_1^{(5)}$, converge as we increase the order of truncated series solution to the BPS equations. We obtain similar results for the single mass case with $\cN=1$ mass deformation.

\subsection{Single mass $\cN=1^*$ model}
This is a special case of $\cN=1$ deformation, where one makes just a single chiral multiplet massive. Namely, $m_1\neq 0 $ and $m_2=m_3=0$. For the supergravity scalars, correspondingly we set $z_1=z_2=z_3=z_4:=z$, $\zt_1=\zt_2=\zt_3=\zt_4:=\zt$, and $\eta_1=\eta^{1/2}$, $\eta_2:=\eta^{3/2}$. Then the action becomes 
\beq
{\cal L}&=-\frac{1}{4}R + 3 \frac{(\partial{\eta})^2}{\eta^2} +2 \frac{ \partial_{\mu} z \partial^{\mu} \bar{z}}{(1-z \bar{z})^2} +{\cal P}(z,\bar{z},\eta) , 
\\
{\cal P} &=-\frac{\eta ^4\left(z^2-1\right) \left(\bar{z}^2-1\right) \left(z^2 \bar{z}^2+z^2-4 z \bar{z}+\bar{z}^2+1\right)}{(z \bar{z}-1)^4} -\frac{2 \left(z^2 \bar{z}^2+z^2+\bar{z}^2+4 z \bar{z}+1\right)}{\eta ^2 (z \bar{z}-1)^2}
\nn\\
&+\frac{\left( z^3+3 z^2 \bar{z}+3 z+\bar{z}\right)\left(\bar{z}^3+3 z \bar{z}^2+3 \bar{z}+z\right) }{2 \eta ^8 (z \bar{z}-1)^4} . 
\eeq
When one analyzes the BPS equations using the UV expansion, 
\beq
\begin{aligned}
 \frac{1}{2}(z+\zt) &=\left(\frac{v}{2}+ \mu \rho\right) e^{-2 \rho}-\left(\frac{v}{8} \mu^2 +\frac{1}{4} \mu^3 \rho\right) e^{-4\rho} +\cdots , 
 \\
 \frac{1}{2}(z-\zt) &=- \frac{\mu}{2} e^{-\rho}+\frac{\mu ^3}{24}  e^{-3\rho}
 + \cdots.
\end{aligned}
\eeq
Again the crucial holographic information is in the relation between $v$ and $\mu$, when we impose IR regularity. Through our series expansion technique, we obtain 
\beq
v=-2 \mu+1.88467 \mu ^3+1.44416 \mu ^5+1.31768 \mu ^7+1.43867 \mu ^9+ \cdots .
\eeq
Note that from our analysis in Sec.\ref{general} the coefficient of $\mu^3$ is in fact
$
\frac{2}{5}+\frac{8 \pi^4}{525} \sim 1.88433
$, so the numerical error is less than 0.02\%.
Recalling how the holographic free energy $F(\mu)$ is related to $v(\mu)$, we have 
\beq
F/ N^2 = 
-0.235584 \mu ^4 -0.120346 \mu ^6 -0.0823552 \mu ^8 -0.0719335 \mu ^{10}+\cdots .
\eeq
which is an improvement over the second equation in Eq.(5.12) of \cite{Bobev:2016nua}.

\subsection{Equal mass $\cN=1^*$ model}
In this case one gives the same non-zero mass to all three chiral multiplets in $\cN=4$ super Yang-Mills theory. On the supergravity side it is implemented by $z_2=z_3=-z_4$, $\zt_2=\zt_3=-\zt_4$, 
$\eta_1=1$, $\eta_2=1$. Then
the action reduces to 
\beq
{\cal L}&=-\frac{1}{4}R + \frac{1}{2}\left(\frac{\partial_\mu z_1 \partial^\mu \zt_1}{(1-z_1 \zt_1)^2}+3\frac{\partial_\mu z_2 \partial^\mu \zt_2}{(1-z_2 \zt_2)^2}\right) +{\cal P},
\\
{\cal P} &=-\frac{3 \left(z_2-1\right) \left(z_2+1\right) \left(z_1 z_2+1\right) \left(\bar z_2-1\right) \left(\bar z_2+1\right) \left(\bar z_1 \bar z_2+1\right)}{\left(z_1 \bar z_1-1\right) \left(z_2 \bar z_2-1\right){}^3}
\nn\\
&+\frac{9 \left(z_2-1\right) \left(z_2+1\right) \left(\bar z_2-1\right) \left(\bar z_2+1\right) \left(z_2+\bar z_1\right) \left(z_1+\bar z_2\right)}{8 \left(z_1 \bar z_1-1\right) \left(z_2 \bar z_2-1\right){}^3}
\nn\\
&-\frac{3 \left(z_2^2 \bar z_2-2 z_1 z_2 \bar z_2+3 z_1 z_2^2+2 z_2-z_1-3 \bar z_2\right) \left(z_2 \bar z_2^2-2 z_2 \bar z_1 \bar z_2-3 z_2+3 \bar z_1 \bar z_2^2+2 \bar z_2-\bar z_1\right)}{8 \left(z_1 \bar z_1-1\right) \left(z_2 \bar z_2-1\right){}^3}.
\eeq
One finds that the UV expansion of the BPS equations leads to the following results,
\beq
\begin{aligned}
\frac{1}{2}(z_1+\bar{z}_1) &=\left(\frac{3}{2}v+ 3 \mu \rho\right) e^{-2 \rho}-\left(\frac{27}{8} v \mu^2 +\frac{27}{4} \mu^3 \rho\right) e^{-4\rho} + \cdots ,
\\ 
\frac{1}{2}(z_1-\bar{z}_1) &=- \frac{3\mu}{2} e^{-\rho}+\frac{9\mu ^3}{8}  e^{-3\rho} 
+ \cdots , 
\\
\frac{1}{2}\left(z_2+\bar{z}_2\right) &=\left(\frac{1}{2}v+ \mu \rho\right) e^{-2 \rho}-\left(\frac{v \mu^2}{8} +\frac{\mu^3}{4}  \rho\right) e^{-4\rho} + \cdots , 
\\
\frac{1}{2}(z_2-\bar{z}_2) &=- \frac{\mu}{2} e^{-\rho}+\frac{\mu ^3}{24}  e^{-3\rho}
+ \cdots .
\end{aligned}
\eeq
Our approximate treatment does not work here as nicely as previous examples. It is because the zero mode part is in general not suppressed, and small errors at lower order propagate to all expansion coefficients at higher orders in $\ep$. We believe it is due to essentially the same difficulty that the authors of \cite{Bobev:2016nua} could extract the coefficients only upto $\mu^4$ in Eq.(5.12) for this case.

\begin{figure}[h]
\includegraphics[width=0.325 \textwidth]{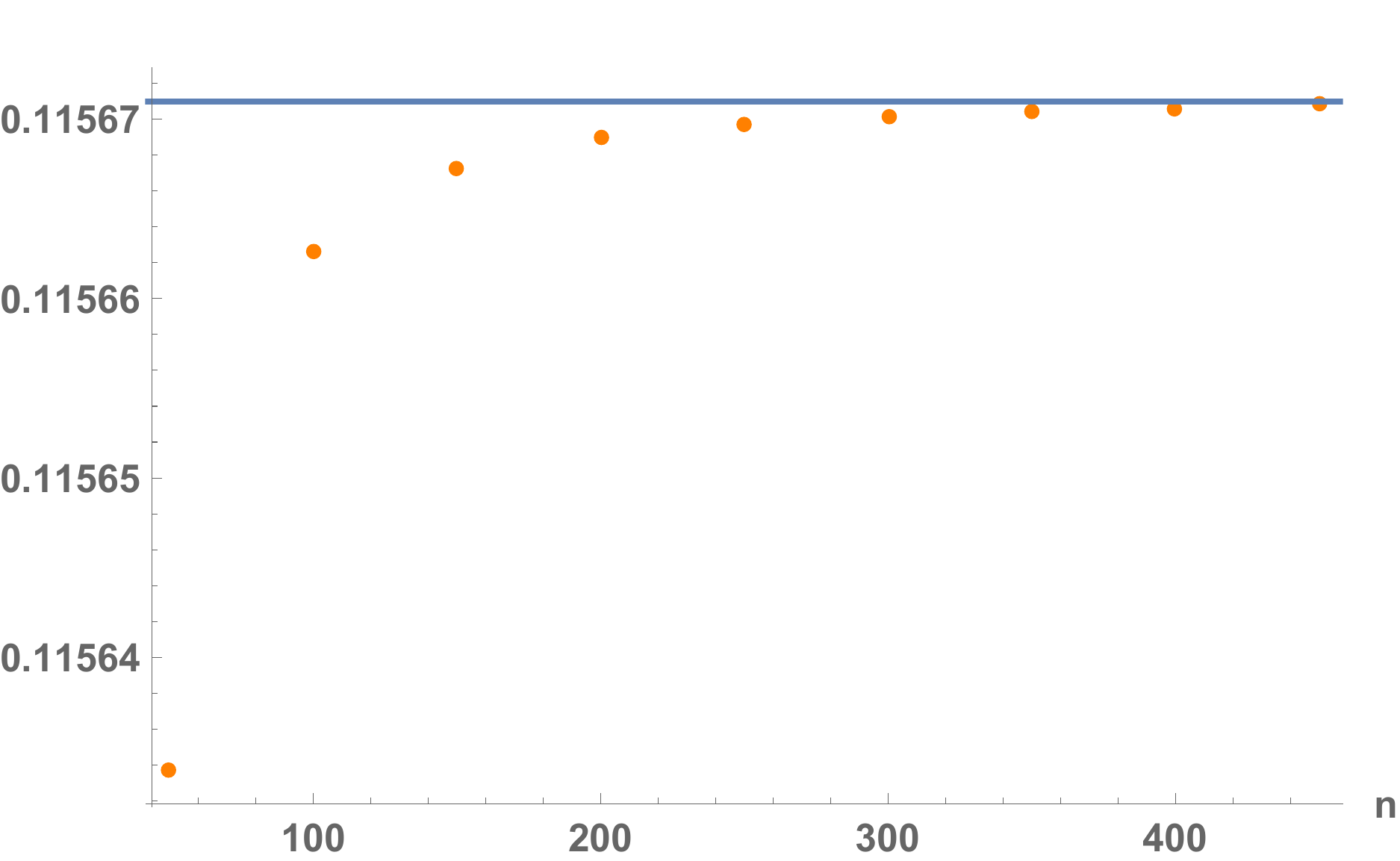} 
\includegraphics[width=0.325 \linewidth]{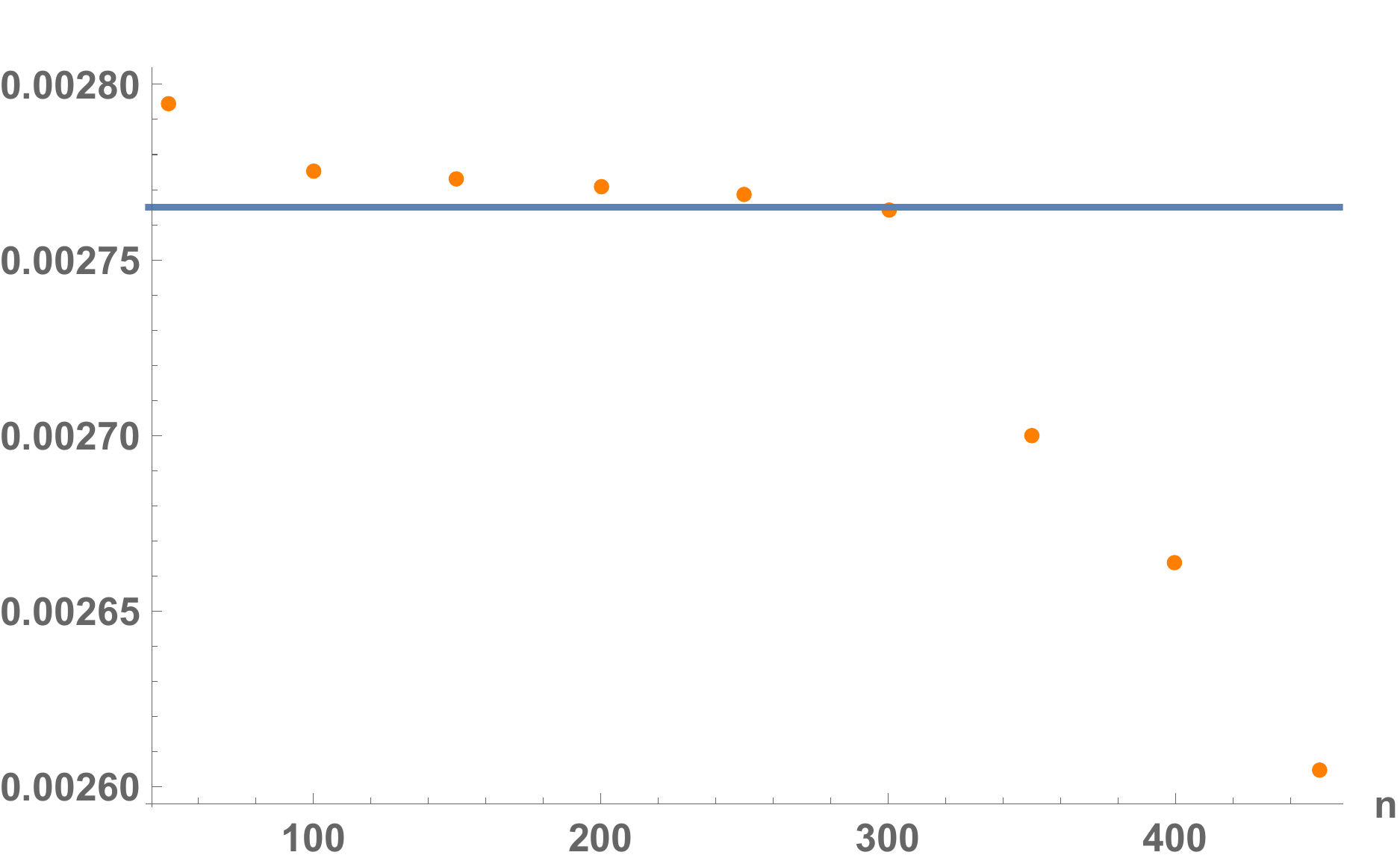} 
\includegraphics[width=0.325 \linewidth]{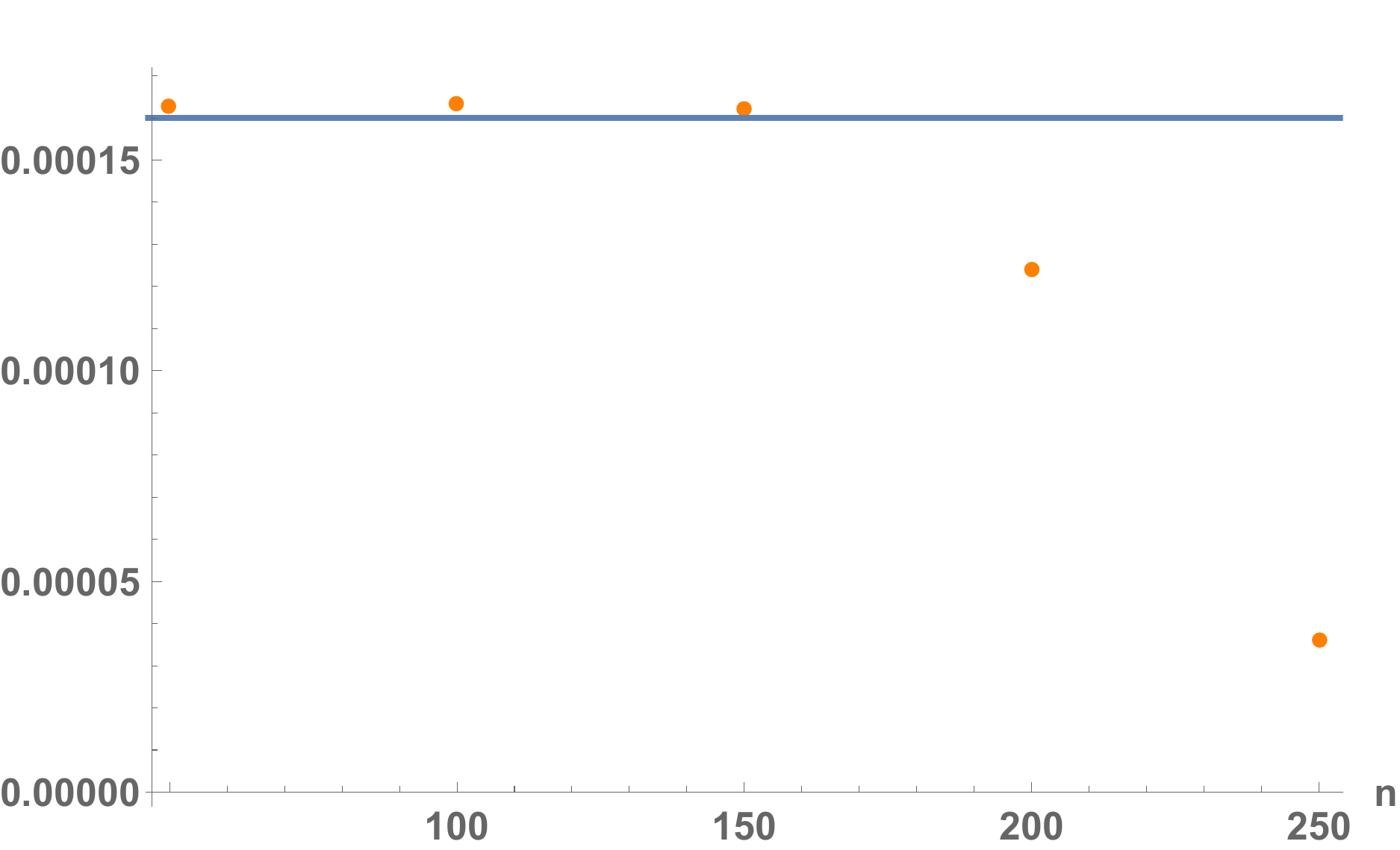} 
  \caption{Plots for the coefficient of $\mu^3,\mu^5,\mu^7$ from the series expansion with Pad\'e approximation, as one increases the truncation length.}
 
  \label{fig2}
\end{figure}

In order to isolate fields with different UV asymptotics, we solve the BPS equations for re-defined functions $y_1\equiv z_1+z_2$, $y_2\equiv  z_1-3z_2$ (and analogously for barred ones). Then the function $v(\mu)$ can be extracted from $y_1$ and $\bar y _1$. Due to the problem of zero modes, it turns out that the UV behavior and equivalently the Taylor coefficients of $v(\mu)$ are best extracted for the truncated series expansion in $r$ with intermediate lengths. We find that the UV limits are stable for the coefficients of $\mu^3$ when we truncate to the range of $300-400$ orders in $r$. For higher orders of perturbation, the limits are best taken within the range of truncation to $150-250$ ($\mu^5$), and $110-150$ ($\mu^7$). See Figure \ref{fig2}. We report, with a truncation upto 150 orders in $r$ and Pad\'e approximation, 
\beq
v=-2 \mu + 0.115668 \mu^3 - 0.00277294 \mu^5 + 0.000162219 \mu^7+\cdots .
\eeq

Then the free energy is 
\beq
F/N^2=-0.0433755 \mu ^4+0.000693235 \mu ^6-0.0000304161 \mu ^8 + \cdots . 
\eeq

Using our results we can also compute the gaugino condensate, $w(\mu)$. Note that this UV parameter vanishes for both ${\cal N}=2^*$ and ${\cal N}=1^*$ single mass deformation. Based on numerical solutions, it was conjectured $w(\mu)=2\mu^3$, {\it precisely}. Our analysis also supports this conjecture, and for instance we have witnessed that the coefficient of $\mu^7$ term in $w(\mu)$ can be suppressed  to less than $10^{-4}$. 
\section{Discussions}
\label{discuss}
In this paper we have re-visited the problem of studying $\cN=1^*$ theories put on $S^4$, using the perturbative prescription recently advocated in \cite{Kim:2019feb}. Presently our goal is to calculate the holographic free energy $F$. In a series expansion form, $F$ is an even function of mass parameters $\mu_i$, starting with $\mu^4$. We have calculated terms up to $\mu^{10}$ for {\it single mass} model, and $\mu^8$ for {\it equal mass} model. Unfortunately however, with our current technology of supersymmetric localization, it is not feasible to calculate the counterparts in gauge field theory since $S^4$ localization needs $\cN=2$ supersymmetry. We hope, one day, our predictions can be confirmed by {\it bona fide} field theory calculations.

Let us summarize our results here to compare with \cite{Bobev:2016nua}, for quick reference. Up to $\cO(\mu^6)$, for symmetry reasons the free energy can be written as 
\beq
F_{S^4}/N^2 &= A_1 (\mu_1^4+\mu_2^4+\mu_3^4)+A_2 (\mu_1^2+\mu_2^2+\mu_3^2)^2
\nn\\
&+B_1 (\mu_1^6+\mu_2^6+\mu_3^6)+B_2 (\mu_1^2+\mu_2^2+\mu_3^2)^3+B_3 \mu_1^2\mu_2^2\mu_3^2
+ \cO(\mu^8).
\eeq
The numerical results of \cite{Bobev:2016nua} give $A_1\approx -0.346$ and $A_2\approx 0.1105$. Their analytic values from our analysis are $A_1 = (105-16\pi^4)/4200 \approx -0.346082$ and $A_2 = (-315+8\pi^4)/4200\approx 0.110541$. For the $\mu^6$-order coefficients, \cite{Bobev:2016nua} gives $B_1\approx -0.146$ and $B_2\approx 0.026$ but failed to calculate $B_3$. Our results give $B_1\approx -0.146573 ,B_2\approx 0.0262266 ,B_3\approx -0.267706$. At this point one may recall the proposal made in Ref. \cite{Gorantis:2017vzz}\footnote{We are grateful to N. Bobev for drawing our attention to this reference after we submitted the first version of this paper to {\tt http://arxiv.org}.}, where the spacetime dimension $d$ is analytically continued to propose some predictions on ${\cal N}=1^*, \, D=4$ free energy. Due to a constraint of their method, they did not make a prediction on $B_1,B_2,B_3$ separately, but conjectured $12B_2-B_3=-1/8$. \footnote{Readers are referred to the 4th equation in Eq.(6.35) of \cite{Gorantis:2017vzz}. Note that $A^{there}_i=-A^{here}_i$ and $C^{there}_i=B^{here}_i$ etc.} We report that our result does {\it not} agree with it, since we have $12B_2-B_3\approx0.58$ instead.

\appendix
\section{Polylogarithm}
Integration of higher components in perturbative treatment involves polylogarithm. We present some useful identities here. They are defined as 
\be
\Li_n(z) = \sum_{k=1}^{\infty} \frac{z^k}{k^n} = z + \frac{z^2}{2^n} + \frac{z^3}{3^n} + \cdots . 
\ee
When analytically continued, $\Li_n(z)$ takes an imaginary value if $\Re(z)>1$. In order to extract the imaginary part, it is useful to recall
\beq
\Li_2(z) &= - \Li_2(1/z) + 2\pi^2\left( \frac{1}{6}  - \frac{i \log z }{2\pi} - \frac{(\log z)^2}{4\pi^2}\right) , 
\\
\Li_3(z) &= + \Li_3(1/z) + \frac{4\pi^3}{3}\left( \frac{\log z}{4\pi} - \frac{3i (\log z)^2}{8\pi^2} - \frac{(\log z)^3}{8\pi^3} \right) , 
\\
\Li_4(z) &= - \Li_4(1/z) + \frac{2\pi^4}{3}\left( \frac{1}{30}+ \frac{(\log z)^2}{4\pi^2} - \frac{i (\log z)^3}{4\pi^3} - \frac{(\log z)^4}{16\pi^4} \right) . 
\eeq
The above formulas are valid for $\Re(z)\ge 1$. We note that the polynomial of $\log z$ in the right-hand-side expressions for $\Li_n(z)$ are from Bernoulli polynomials of $n$-th order.

One finds that the explicit integrations can be expressed in terms of even and odd parts of polylogarithms,
\beq
\pi _s (z):=\sum_{n=1}^{\infty} \frac{z^{2n}}{(2n)^s} , \quad\chi_s (z):=\sum_{n=1}^{\infty} \frac{z^{2n-1}}{(2n-1)^s} .
\eeq

\section{Explicit form of some perturbative solutions and UV asymptotics}
The UV behavior of first order solutions $u_1,u_2$ are, in terms of 
$\rho=2 \tanh ^{-1} (r)$, 
\beq
u_1&=+2 e^{-\rho }-4e^{-2 \rho } \left( \rho -1\right)-2e^{-3 \rho } \left(4 \rho -3\right)-8e^{-4 \rho } \left(2\rho -1\right)+{\cal  O}(\rho e^{-5\rho}),
\\
u_2&=-2 e^{-\rho }-4e^{-2 \rho } \left( \rho -1\right)+2e^{-3 \rho } \left(4 \rho -3\right)-8e^{-4 \rho } \left(2\rho -1\right)+{\cal  O}(\rho e^{-5\rho}).
\eeq
Using the above, the leading order solution for $z_i$ are given as follows. 
\beq
z_1&=s-\frac{1}{2} (1-s^2) \left(\mu _1+\mu _2+\mu _3\right)e^{-\rho}
\nn
\\
&-\frac{1}{4} \left(1-s^2\right) \left(\left(\mu _1+\mu _2+\mu _3\right){}^2 s-2( v_1+ v_2+ v_3)-4\left(\mu _1+\mu _2+\mu _3\right) \rho\right)e^{-2\rho},
\\
z_2&=-s-\frac{1}{2} (1-s^2) \left(\mu _1-\mu _2+\mu _3\right)e^{-\rho}
\nn\\
& +\frac{1}{4} \left(1-s^2\right) \left(\left(\mu _1-\mu _2+\mu _3\right){}^2 s+2( v_1- v_2+ v_3)+4(\mu _1-\mu _2+\mu _3)\rho\right)e^{-2\rho},
\\
z_3&=-s-\frac{1}{2} (1-s^2) \left(\mu _1+\mu _2-\mu _3\right)e^{-\rho}
\nn\\
&+\frac{1}{4} \left(1-s^2\right) \left(\left(\mu _1+\mu _2-\mu _3\right){}^2 s+2 (v_1+ v_2- v_3)+4(\mu _1+\mu _2-\mu _3)\rho\right)e^{-2\rho},
\\
z_4&=s-\frac{1}{2}(1-s^2) \left(\mu _1-\mu _2-\mu _3\right)e^{-\rho}
\nn\\
&-\frac{1}{4} \left(1-s^2\right) \left(\left(\mu _1-\mu _2-\mu _3\right){}^2 s-2( v_1- v_2- v_3)-4(\mu _1-\mu _2-\mu _3)\rho \right)e^{-2\rho}.
\label{muv}
\eeq
This UV expansion is as given by \cite{Bobev:2018hbq}. The relations between integration constants are
\beq
\begin{aligned}
c_1 \epsilon&=\frac{1}{8} (v_1^{(1)} +v_2^{(1)}+v_3^{(1)}), & c_2 \epsilon & = \frac{1}{8}(v_1^{(1)} -v_2^{(1)}+v_3^{(1)}),
\\
c_3 \epsilon& = \frac{1}{8} (v_1^{(1)} +v_2^{(1)}-v_3^{(1)}), & c_4 \epsilon & = \frac{1}{8}(v_1^{(1)} -v_2^{(1)}-v_3^{(1)}).
\end{aligned}
\eeq

We now present the correction to warp factor at $\cO(\ep^2)$.
\beq
{\cal A}^{(2)} &= \frac{3}{4}\left( \chi _3\left(\frac{1-r}{1+r}\right)+2\tanh ^{-1}(r)\chi _2\left(\frac{1-r}{1+r}\right)-\tanh ^{-1}(r)^2\log (r)\right)
\nn\\
&+\frac{r^6+33 r^4-33 r^2-1}{96 r^2 \left(1+r^2\right)}-\frac{\left(r^6-21 r^4+21 r^2-1\right) \tanh ^{-1}(r)}{48 r^3}
\nn\\
&+\frac{\left(r^{10}+r^8+64 r^6-64 r^4-r^2-1\right) \tanh ^{-1}(r)^2}{96 r^4 \left(1+r^2\right)}.
\eeq
The scalar fields $\eta_1,\eta_2$ at $\cO(\ep^2)$ are given as 
\beq
\eta _{1}^{(2)}(r)=&-\frac{\mu _1^2+\mu _2^2-2 \mu _3^2}{24 }\frac{\left(1-r^2\right)^2}{4 r^2}\left(1-2\frac{\left(1+r^2\right)}{ r}  \tanh ^{-1}(r)+\frac{\left(1-r^2\right)^2}{ r^2} \tanh ^{-1}(r)^2\right),
\nn\\
\eta _{2}^{(2)}(r)=&-\frac{\mu _1^2-\mu _2^2}{8 }\frac{\left(1-r^2\right)^2}{4 r^2}\left(1-2\frac{\left(1+r^2\right)}{ r}  \tanh ^{-1}(r)+\frac{\left(1-r^2\right)^2}{ r^2} \tanh ^{-1}(r)^2\right).\nn
\eeq

\acknowledgments
We are grateful to N. Bobev for his comments on the first version of this paper, in particular for drawing our attention to holographic gaugino condensate $w(\mu)$, and also to Ref. \cite{Gorantis:2017vzz}. This research was supported by the National Research Foundation of Korea (NRF) grant 2018R1D1A1B07045414. Se-Jin Kim was supported by the CERN-Korea graduate student visiting program supported by NRF.

\bibliographystyle{JHEP}
\bibliography{onestar}

\end{document}